# Dark gamma-ray bursts: possible role of multiphoton processes

Mark E. Perel'man[1)]

(*Racah Institute of Physics, Hebrew University, Givat Ram, Jerusalem, Israel*)

**ABSTRACT** The absence of optical afterglow at some gamma-ray bursts (so called dark bursts) requires analyses of physical features of this phenomenon. It is shown that such singularity can be connected with multiphoton processes of frequencies summation in the Rayleigh- Jeans part of spectra, their pumping into higher frequencies. It can be registered most probably on young objects with still thin plasma coating, without further thermalization, i.e. soon after a prompt beginning of the explosive activity.

========================================

Gamma-ray bursts represent the plurality of "hot" problems for astrophysics and fundamental physics (e.g. [1]). Among them the phenomenon of so called "dark" gamma-ray bursts, events without detected optical afterglow or with detected optical flux significantly fainter than expected from the observed X-ray afterglow, is of great interest and attracts a big attention. In the review [2] several proposed mechanisms of decreasing the visual part of spectra are cited and considered, among them are high redshift, dust along the line of sight and some more exotic suggestions of radiation sources features, but none of them gives satisfactory explanation of the phenomenon.

Let us consider the mechanism that obligatory must take part in formation or reformation of spectra of bright sources: the multiphoton processes at passage of intensive photon flux through clouds of electrons and ions. Some of them are examined in the general form in [3], Section7, earlier they were announced in [4].

The most significant in this connection can be processes of photons energy summation. They are widely investigated in laser physics: generation of higher harmonics (HHG, e.g. [5]), especially onto single electrons, can be considered by an analogy with a multiphoton Compton scattering; the HHGs at the accelerator energy had been also investigated [6]. But the dependence on densities of photons fluxes with different frequencies represents some new problems.

In the frame of quantum electrodynamics (QED) there is not a principal difference between processes with photons of equal frequencies, the most usual in laser physics (cf. nevertheless [7]), and with photons of different frequencies: all difficulties are of computing character only. Therefore we shall consider these processes, as distinct from [3], by methods of quantum kinetics in application to the black body radiation and to the synchrotron source also.

---

[1]). E-mails: m.e.perelman@gmail.com; mark_perelman@mail.ru

* * *

Let us consider the multiphoton process:

$$\gamma(\omega_1) + \cdots + \gamma(\omega_M) + e \to e + \gamma(\Omega) \tag{1}$$

on free or bound electron with $\Omega \lesssim \sum_1^M \omega_i$ that can be described as an inverse multiphoton Compton process.

If photon flux corresponds to the black body radiated on with temperature T, this process can be executed in the Rayleigh- Jeans (R-J) region with big density of photons.

The number of photons of definite frequency emitted into the volume V is equal in this approximation to

$$dN(\omega) = \frac{V}{\pi^2 c^3} \frac{\omega d\omega}{e^{\hbar\omega/\kappa T}-1} \to \frac{V \kappa T \omega d\omega}{\pi^2 c^3 \hbar}. \tag{2}$$

The complete number of photons emitted by the black body $N \approx \frac{1}{4}(\kappa T/\hbar c)^3 V$ and the relative probability of low frequencies R-J quanta

$$n(\omega) = N(\omega)/N \sim \frac{2}{\pi^2}\left(\frac{\hbar\omega}{\kappa T}\right)^2. \tag{3}$$

Probabilities of (1) should be calculated via effective magnitudes that depend on temperature. Let's begin their estimation.

We shall begin with averaging of the free path length $\ell = 1/\rho\sigma_T$, where in accordance with the optical theorem of QED $\sigma_T = 4\pi c r_0/\omega$ is the total cross-section of interaction (virtual absorption) of low frequencies photons, $r_0$ is the classical radius of electron. Its averaging over the interval $[0, \varpi]$ can be determined as

$$\ell_{\text{eff}}(\varpi) = \frac{1}{\varpi}\int_0^\varpi \frac{1}{\rho\sigma_T(\omega)} n(\omega) d\omega = \frac{\varpi}{2\pi^3 \rho c r_0}\left(\frac{\hbar\varpi}{\kappa T}\right)^2. \tag{4}$$

This magnitude allows an estimation of the averaged total cross-section:

$$\sigma_{\text{eff}}(\varpi) = \frac{1}{\rho \ell_{\text{eff}}(\varpi)} = \frac{2\pi^3 c r_0}{\varpi}\left(\frac{\kappa T}{\hbar\varpi}\right)^2. \tag{5}$$

The duration of the process (1) is determinable as $\tau \sim 1/\Omega$ [8] and the flux of virtually absorbed photons can be estimated as

$$J_0 = \frac{\hbar\varpi}{\sigma_{\text{eff}}(\varpi)\tau(\Omega)}. \tag{6}$$

We can assume, for simplicity, that the limiting frequency of summation process $\varpi = \Omega/2$, this choice has not principal sense. Processes of (1) type have significant magnitude if

$$J(\varpi) \geq J_0(\varpi) \tag{7}$$

with the R-J flux $J(\varpi) = \kappa T \varpi^3 / \pi^2 c^2$. This condition leads to the main estimation of maximal frequencies achievable by summation process:

$$\left(\frac{\hbar\Omega}{\kappa T}\right)^2 \leq 4\pi \frac{r_0}{c} \frac{\kappa T}{\hbar} \tag{8}$$

or

$$\Omega \leq 6 \cdot 10^4 T^{3/2}. \tag{9}$$

The estimation shows that, for example, at $\kappa T = 1$ MeV, i.e. T~$10^{10}$ K into higher parts of spectra will be throw quanta with $\hbar\Omega \leq 420$ keV and, correspondingly, "low frequency" part of spectrum will be impoverished. But for the external surfaces of usual stars with $\kappa T$~1 eV the limiting frequency $\hbar\Omega \leq 4.2 \cdot 10^{-4}$ eV, i.e. the considered effect can be observable in the far IR and in microwaves.

\* \* \*

The offered method can be applied to other sources of radiation also.

So, for the synchrotron radiation instead of (2) must be written

$$dN_{SR}(\omega) = C \cdot H_\perp \omega^a e^{-b\omega} d\omega. \tag{10}$$

Thus, independently from the magnitude of **H**,

$$n_{SR}(\omega) = N_{SR}(\omega)/N_{SR} \sim \omega^a e^{-b\omega}$$

and for the most cases can be taken $b = 0$ and $a \sim \frac{1}{3} \div \frac{1}{2}$.

Hence, in the complete analogue with considerations executed above ($b = 0$, for brevity)

$$\ell_{\text{eff}}(\varpi) = \frac{\varpi^{a+1}}{4\pi\rho c r_0(a+2)}; \qquad \sigma_{\text{eff}}(\varpi) = \frac{1}{\rho\ell_{\text{eff}}(\varpi)} = \frac{4\pi c r_0(a+2)}{\varpi^{a+1}}$$

$$J_0 = \frac{\hbar\varpi}{\sigma_{\text{eff}}(\varpi)\tau(\Omega)} = \frac{\hbar\varpi^{a+3}}{2\pi c r_0(a+2)}. \tag{11}$$

As $J_{SR}(\omega) = c\hbar\omega N_{SR}(\omega)$, from the condition of saturation follows the analog of (9):

$$\Omega \leq C \frac{a+2}{a+1} 4\pi c^2 r_0 \cdot H_\perp, \tag{12}$$

which must be considered for concrete types of stars and bursts.

\* \* \*

Against the considered effect will work thermalization of high energy photons and impossibility of photons summations on account of scattering of electrons during their excitation. If the mean velocity of electrons $v = \sqrt{\kappa T/3m}$, the nearest frequencies that can take part in the process would be estimated via relation between free path lengths:

$$v/\omega_{min} \gg \ell_{eff}(\omega_{min}).$$

On the other hand a realization of the effect depends on possibility of avoiding thermalization in the upper plasma layers: its mean depth L must be not very bigger the free path length relative to photon-electron scattering.

This condition means that "dark" gamma-bursts can be observable at comparatively young bursts with underdeveloped plasma atmospheres.

In conclusion we can state that the phenomenon of "dark" gamma-bursts is caused by multiphoton summation of frequencies at initial stages of their evolutions independently from the concrete mechanism of burst.